\documentclass[twoside]{LCWS11}
\usepackage[latin1]{inputenc}
\usepackage[dvips]{graphicx,epsfig,color}
\usepackage{wrapfig,rotating,multirow}
\usepackage{amssymb,amsmath,array}
\usepackage{cite}
\begin{document}
\title{
 Development of CMOS Pixel Sensors fully adapted to the ILD Vertex Detector Requirements} 
\author{Marc Winter, 
J\'er\^ome Baudot, Auguste Besson, Gilles Claus, Andre\"{i} Dorokhov, 
Mathieu Goffe, \\ 
Christine Hu-Guo, Fr\'ed\'eric Morel, Isabelle Valin, Georgios Voutsinas and Liang Zhang   
\vspace{.3cm}\\
Institut Pluridisciplinaire Hubert Curien (IPHC) - Universit\'e de Strasbourg \\
23 rue du loess, BP-28, 67037 Strasbourg Cedex 2 - France
\vspace{.1cm}\\
}

\maketitle

\begin{abstract}
CMOS Pixel Sensors are making steady progress towards the specifications
of the ILD vertex detector. Recent developments are summarised, which 
show that these devices are close to comply with all major requirements,
in particular the read-out speed needed to cope with the beam related 
background. This achievement is grounded on the double-sided ladder 
concept, which allows combining signals generated by a single particle 
in two different sensors, one devoted to spatial resolution and the 
other to time stamp, both assembled on the same mechanical support. 
The status of the development is overviewed as well as the plans to 
finalise it using an advanced CMOS process.
\end{abstract}

\section{Introduction}

   The ILC physics programme encompasses numerous studies relying on high 
precision flavour tagging, including high quality charmed meson and 
tau lepton identification. This objective translates into the necessity of
a very precise vertex detector, to be equipped with very granular and thin 
pixel sensors. Taking advantage of the ILC running conditions , which are 
much less demanding than those at the LHC, physics driven specifications 
such as spatial resolution can be privileged at the expense of read-out 
speed or radiation tolerance. 

   CMOS Pixel Sensors (CPS) constitute a category of Monolithic Active 
Pixel Sensors (MAPS) offering attractive features for such requirements.
They easily match the targeted granularity and material budget, and do 
not necessitate a cooling system adding substantial material budget inside 
the fiducial volume of the detector. On the other hand, the hit rate 
generated by the beam related background in the innermost layer of 
the detector sets target values of the read-out speed which are not 
straightforward to achieve, given the size, and thus the number, of the
pixels needed to match the spatial resolution. Speed is therefore a
major driving parametre of the development of CPS since a few years.

\section{State-of-the-art CPS}

\subsection{Sensor short description}

   The sensors MIMOSA-26~\cite{M26} and -28~\cite{M28} sensors may 
be considered as the state-of-the-art of the CPS technology for 
charged particle tracking. They were realised to equip the EUDET 
beam telescope~\cite{eudet} and 
the new STAR vertex detector~\cite{PXL}, respectively. Their pixels 
are grouped in columns read out in parallel and terminated with an 
offset compensated discriminator. Inside each column, the pixels 
are read out sequentially with a typical read-out time of 
$\lesssim$ 200~ns per pixel. This so-called {\it rolling 
shutter} mode exhibits the great advantage of limiting the 
power consumption of the whole pixel array to the amount needed
to operate a single row. MIMOSA-26 and -28 exhibit power consumptions
of about 250 and 150~mW/cm$^2$ respectively, the spread between both 
values reflecting the geometry differencies of the two sensors.

  The pixel pitch is around 20 $\mu m$ for both sensors. Each 
pixel incorporates a micro-circuit allowing for correlated 
double-sampling for the purpose of average pixel noise subtraction, 
and a pre-amplification stage mitigating the impact of the 
noise sources of the signal processing chain downstream of the 
pixels. The discriminator thresholds, as well 
as the settings of most of the sensors' steering parametres, are 
remotely programmable through a JTAG circuitry integrated on the 
sensor. A zero-suppression circuit integrated in the chip periphery 
transforms the signals in excess of the discriminator thresholds 
into hit pixel addresses which are buffered in integrated SRAMs 
before being transmitted to the outside world. 

\section{Achieved detection performances}

\begin{wrapfigure}{r}{0.5\columnwidth}
\centerline{\includegraphics[width=0.53\columnwidth]{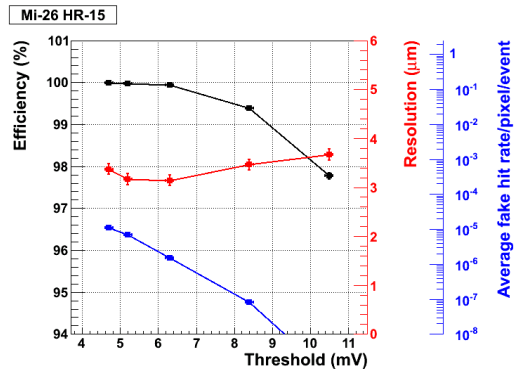}}
\caption{MIMOSA-26 beam test results obtained at the CERN-SPS 
with $\sim$ 10$^2$~GeV charged particles. The detection efficiency 
(in black), the fake hit rate (in blue) and the single point 
resolution (in red) are shown for various values of the 
discriminator thresholds.}
\label{Fig:M26}
\end{wrapfigure}

   The detection performances of severals tens of sensors were 
assessed with minimum ionising particle beams at CERN and DESY.
Numerous measurements were performed with 50~$\mu m$ thin sensors, 
at various operation temperatures and after exposures to various 
integrated 
ionising and non-ionising radiation doses. Figure \ref{Fig:M26} 
shows typical values of the detection efficiency, pixel fake hit 
rate (due to noise fluctuations above threshold) and single point 
resolution obtained with a MIMOSA-26 sensor at a temperature of 
about 20$^{\circ}$C for various values of the discriminator 
thresholds.

    One observes that the detection efficiency stays close to 100~\% 
for threshold values high enough to keep the fake hit rate at a 
negligible level (e.g. $\lesssim$ 10$^{-4}$). The single point 
resolution is only slightly in excess of the ILD specification 
of 3~$\mu m$~\cite{LoI}. It results from an impact position 
reconstruction based on the centre of gravity of the positions
the few pixels composing a cluster. The value 
obtained is well below the digital resolution reflecting the 
18.4~$\mu m$ pixel pitch of MIMOSA-26 (i.e. 5.3~$\mu m$) 
despite the binary charge encoding. 
It follows that a pitch of $\lesssim$ 17~$\mu m$
would allow complying with the 3~$\mu m$ spatial resolution 
required for the ILD vertex detector. It was actually checked 
that the discriminators ending the columns would fit within 
$<$17~$\mu m$ wide columns. These performances hold for 
irradiated sensors (e.g. MIMOSA-28 was validated for 150 kRad 
and 3$\times$10$^{12}$n$_{eq}$/cm$^2$ at a temperature of 
30-35$^{\circ}$C).

   The 576 pixels composing the 1152, 10.5 mm long, columns 
of MIMOSA-26 are read out from one side in about 100~$\mu s$. 
The read-out would be twice faster, i.e. $\sim$ 50~$\mu s$ 
short, in case of a double-sided read-out architecture where 
each column is split in two opposite halves. As explained 
later in this paper, this value is expected to be appropriate 
for the innermost layer of the ILD vertex detector, which is 
exposed to the highest particle rate. The twice higher power
consumption resulting from the double-sided read-out is still
expected to be compatible with low mass air cooling.

\section{Concept developed for the ILD vertex detector}

   As indicated in the previous section, the compliance of 
CPS with the single point resolution and the material budget 
specifications of the ILD vertex detector are not questionable.
The measured radiation tolerance of MIMOSA-26 and -28 is also 
expected to be sufficient for the running conditions foreseen. 
The remaining questions are whether the read-out speed can 
accommodate the hit rate generated by the beam related
background, and whether the power consumption is compatible 
with a non-disturbing cooling mean such as air flow. 

   The particle rate is dominated by beamstrahlung electrons, 
and decreases rapidly when moving away from the interaction 
region~\cite{LoI}. For instance, a detector geometry based on 
three cylindrical double-sided layers featuring average radii 
of 17, 38 and 59 mm, faces a hit density varying by one ordre 
of magnitude from one layer to the next. On the other hand, 
the innermost layer, which is by far the most exposed to 
beamstrahlung background, is also the smallest one, standing 
for only about 10~\% of the total detector surface. These 
features are exploited in the concept developed here, together 
with the design flexibility and the moderate cost of the CPS 
technology. 

   Three different sensors are foreseen, each optimised for 
a different balance between the single point resolution, the
read-out speed and the power consumption. The $\lesssim$ 
3~$\mu m$ resolution and the fast read-out needed for the 
innermost layer are provided by two different sensors, 
implemented on the two faces of the ladders equipping the 
layer. 

\begin{wrapfigure}{r}{0.5\columnwidth}
\centerline{\includegraphics[width=0.5\columnwidth]{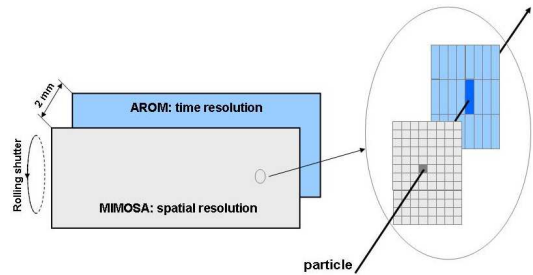}}
\caption{Schematic view of the combination of AROM (elongated pixels)
and MIMOSA (square pixels) sensors equipping a 2~mm thick double-sided 
ladder.}
\label{Fig:2sided}
\end{wrapfigure}

    One sensor, called MIMOSA-in, provides the spatial 
resolution with 17$\times$17~$\mu m^2$ pixels read out in 
$\sim$ 50~$\mu s$. The other sensor, called AROM\footnote{
{\bf AROM} stands for {\bf A}ccelerated {\bf R}ead-{\bf O}ut 
{\bf M}{\sc imosa} sensor.}, features 17$\times$85~$\mu m^2$ 
pixels elongated in the direction of the 
columns. It is therefore read out in $\sim$ 10~$\mu s$, the 
columns being composed of $\sim$ 5 times less pixels. Based 
on beam test results of a sensor prototype featuring 
18.4$\times$73.2~$\mu m^2$ pixels, its spatial resolution 
is expected to be $\lesssim$ 6~$\mu m$ in both directions 
with staggered pixels~\cite{M22AHR}. 

   Particles traversing the layer will thus get assigned 
a spatial resolution of $\lesssim$ 3~$\mu m$ combined 
with a $\sim$ 10~$\mu s$ time stamp from their two, $\sim$ 
2~mm apart, impacts. The approach is illustrated in 
Figure\ref{Fig:2sided}. The ambitionned ladder total material 
budget amounts to $\lesssim$ 0.3~\% of radiation length. 
Details on the ladder design and development may be found 
in~\cite{JB}.

   The outer layers, which are less demanding in terms of 
spatial resolution and read-out speed, are foreseen to be 
equipped with a sensor consuming at least 3 times less power, 
called MIMOSA-out. It is supposed to provide a single point 
resolution of $\sim$ 4~$\mu m$ and a read-out time of $\sim$ 
100~$\mu s$. The pixels are 34$\times$34~$\mu m^2$ large, 
i.e. 4 times larger than the pixels of MIMOSA-in. Despite 
the sizeable pitch, a good spatial resolution is expected 
from replacing the discriminators ending the columns with 
$\lesssim$ 4-bit ADCs incorporating a discriminator stage. 
The zero-suppression circuitry of MIMOSA-in can easily be
adapted to this change of the charge encoding.

\begin{wraptable}{r}{0.5\columnwidth}
\renewcommand{\arraystretch}{1.2}
\centerline{\begin{tabular}{|l|c|c|c|}
\hline
\bf Layer & {\bf Specs} & {\sc Mimosa} & {\sc Arom} \\[2mm]
\hline
\multirow{2}{*}{\bf Inner} & $\lesssim$ 3~$\mu m$ &
$\lesssim$ 3~$\mu m$ & $\lesssim$ 6~$\mu m$ \\[1mm]
& 25-50~$\mu s$ &   50~$\mu s$   &   10~$\mu s$ \\[1mm]
\hline
\multirow{2}{*}{\bf Outer} & $\lesssim$ 5~$\mu m$ &
$\lesssim$ 4~$\mu m$ & --  \\[1mm]
& 100-200~$\mu s$ &  100~$\mu s$  & --  \\
\hline
\end{tabular}}
\caption{ILD vertex detector specifications for the single point
and time resolutions in the innermost and outer layers. The 
expected performances of the 3 sensors (MIMOSA-in and -out, AROM)
proposed to equip the detector are shown in comparison.}
\label{tab:sensors}
\end{wraptable}

   The expected spatial and temporal resolutions 
of the three sensors are summarised on 
Table~\ref{tab:sensors}, where they are 
compared to the detector specifications. 
The latter are clearly within reach of the 
CPS proposed, which are not expected to 
face any challenge given the performances 
already achieved with existing sensors. 
The finalisation of the development is 
well under way and should converge within 
a few years, as explained in the next section.

\section{Sensor finalisation plans}

   The development of the three sensors does not require 
the same effort for all of them. While MIMOSA-in and AROM 
are relatively straightforward to realise, MIMOSA-out still 
requires establishing the operation of fast ADCs ending the 
columns (and the corresponding zero-suppression micro-circuitry).

\subsection{Validation of the sensor architectures}

   The validation of the architectures of MIMOSA-in and AROM 
motivated the realisation of the MIMOSA-30 prototype, which 
combines the two designs. It is split in two halves, each made
of 128 columns ended with a discriminator. One half features 
columns of 256, 16$\times$16~$\mu m^2$ large, pixels standing
for a section of MIMOSA-in. The other half differs by the pixel
size and number, i.e. the columns contain 64 elongated pixels 
of 16$\times$80~$\mu m^2$, and stands for a section of AROM.
The read-out times of the two halves amount to the nominal 
values of MIMOSA-in and AROM, i.e. about 50 and 10~$\mu s$ 
respectively and the pixel dimensions were chosen to safely 
comply with the ILD spatial resolution requirement. The chip 
was fabricated at the end of 2011 and its performances are 
foreseen to be assessed at the CERN-SPS in Spring or Summer 
2012.     

   The MIMOSA-out architecture, with its integrated ADCs, is
being prototyped with the MIMOSA-31 chip, also fabricated by 
the end of 2011. The ADC design resembles the Successive 
Approximation Register architecture. It features a variable
charge encoding granularity, ranging from a maximum of 4 bits 
for signals of small magnitude to only 2 bits for large signals. 
Earlier prototypes allowed to check that a rough encoding of 
the amplitude delivered by those pixels in a cluster having
collected the largest charges does not degrade the resolution
on the reconstructed impact position. The precise performances 
of the sensor will be investigated at the CERN-SPS in Autumn 
2012.

\subsection{Translation in an advanced fabrication process}
   
    Most of the sensors realised up to now, including MIMOSA-26 
and -28 as well as MIMOSA-30 and -31 (see previous sub-section), 
were manufactured in a 0.35~$\mu m$ CMOS process. The latter 
is far from relying on fabrication parametres allowing to
approach the real potential of the CPS technology. For instance,
the number of metalisation layers, limited to 4, complicates
substantialy the integration of the ADCs in MIMOSA-out. The next
steps of the development will therefore rely on a 0.18~$\mu m$
process, which features several improvements with respect to the
0.35~$\mu m$ process. Besides the improved ionising radiation
tolerance consecutive to the thinner gate oxyde, the process 
offers 6-7 metalisation layers and deep p-wells allowing 
to use both types of transistors inside the pixels without 
substantial parasitic charge collection by the n-type zones 
hosting {\sc p-mos} transistors.     

    The first prototype exploring this new technology was 
fabricated during the last quarter of 2011. It includes various 
pixel designs exploring different charge sensing and in-pixel 
amplification options. It will be followed in 2012 by prototypes 
reproducing the architectures of MIMOSA-30 and of the 
as the zero-suppression circuitry necessary to complete the
final sensor design. Various architectures will actually be 
explored. Some of them are intended to squeeze the AROM read-out 
time to a few $\mu s$. One motivation for this short integration 
time is the necessity to accommodate the higher beamstrahlung 
background expected when the ILC will be running around 1~TeV 
collision energy. The short integration time is also relevant 
for the running at 500~GeV, for instance to mitigate the 
need for a dedicated soleno\"{i}d installed inside the beam 
pipe to protect the vertex detector from low energy beam 
background electrons backscattered from beam elements located 
near the interaction point. If required, it would also allow 
for a higher granularity of the MIMOSA-in sensor, which would
slow down its read-out speed.

   Detailed estimates of the total power of the detector were
performed with various sensor configurations and assumptions 
on the AROM sensor read-out speed. These computations indicate 
an instantaneous consumption in the ordre of 0.5~kW. Assuming 
a conservative sensor duty cycle of 2~\%, while the machine 
duty cycle is about 0.5~\%, the average power consumption 
of the whole detector would amount to $\sim$ 10~W. This value
is expected to be modest enough to be compatible with an air 
flow based cooling system.

\section{Summary}

   Three different CMOS pixel sensors are being developed 
for the ILD vertex detector, adapted to an original concept 
privileging the spatial and temporal resolutions in the 
innermost layer, and minimising the power consumption in the
outer layers. Their development is well advanced, translating 
into the perspective of fabricating full scale prototypes
complying with all detector specifications within the few 
coming years. In particular, the fast read-out imposed on 
the sensors equipping the innermost layer of the detector is 
nearly achieved, with the perspective of further improvements
reducing the integration time to a few microseconds only, 
well beyond the requirements at a 500~GeV collision energy, 
but well suited to the machine operation near 1~TeV.     
 
   The finalisation of the development relies in particular 
on a 0.18~$\mu m$ CMOS process currently under study. This 
R\&D programme is pursued in synergy with the ALICE-ITS 
upgrade effort~\cite{ITS} as well as in the perspective of 
the upcoming CBM experiment at FAIR~\cite{CBM}, for which 
the fast variants of the AROM sensor are particularly 
attractive.

\newpage

\begin{footnotesize}

\end{footnotesize}
      

\begin{thebibliography}{99}
\bibitem{M26} A.~Dorokhov {\it et~al.}, Nucl. Inst. \& Meth. {\bf A650}, 174 (2011).
\bibitem{M28} I.~Valin {\it et~al.}, TWEPP-2011 Conf. Proc. (2011), to be pub. in JINST. 
\bibitem{eudet} A.~Besson {\it et~al.}, arXiv:1201.4657v1 [physics.ins-det] (2011). 
\bibitem{PXL} L.~Greiner {\it et~al.}, Nucl. Inst. \& Meth. {\bf A650}, 68 (2011).  
\bibitem{LoI} ILD Concept Group, ILD Letter of Intent, arXiv:1006.3396v1 [hep-ex] (2010)
\bibitem{M22AHR} A.~Besson {\it et~al.}, ICHEP-2011 Conf. Proc. (2012). 
\bibitem{JB} J.~Baudot {\it et~al.}, these proceedings. 
\bibitem{ITS} ALICE collaboration, ALICE-ITS upgrade Conceptual Design Report, 
              (2012). 
\bibitem{CBM} Ch.~Schrader {\it et~al.}, XLVIII Int. Winter Meet. 
on Nucl. Phys., Conf. Proc. (2010).
\end{thebibliography}
\end{document}